\documentclass[journal]{IEEEtran}
\usepackage{mathrsfs}
\usepackage{cite}
\usepackage{amsmath}
\usepackage{amssymb}
\usepackage{stfloats}
\usepackage{graphicx}
\usepackage{setspace}
\usepackage{xcolor}

\ifCLASSOPTIONcompsoc
\usepackage[tight,normalsize,sf,SF]{subfigure}
\else
\usepackage[tight,footnotesize]{subfigure}
\fi

\begin{document}
\title{Truncated Beam Sweeping for Spatial Covariance Matrix Reconstruction in Hybrid Massive MIMO}
\author{Yinsheng~Liu, Hongtao~Duan, and Xi Liao.
%\thanks{Copyright (c) 2015 IEEE. Personal use of this material is permitted. However, permission to use this material for any other purposes must be obtained from the IEEE by sending a request to pubs-permissions@ieee.org.}
%\thanks{Zihao Fu is with Electromagnetic Environment Effect Laboratory, Beijing Institute of Radio Metrology and Measurement, Beijing 100854, China and Information School, Communication University of China, Beijing 100024, China.}
\thanks{Yinsheng Liu is with State Key Laboratory of Rail Traffic Control and Safety, Beijing Jiaotong University, Beijing 100044, China, and Frontiers Science Center for Smart High-spped Railway System (e-mail: ys.liu@bjtu.edu.cn).}
\thanks{Hongtao Duan is with Beijing radio monitoring station of State Radio Monitoring Center (SRMC), Beijing 100037, China.}
\thanks{Xi Liao is with Chongqing Key Laboratory of Complex Environmental Communications, School of Communication and Information Engineering, Chongqing University of Posts and Telecommunications, Chongqing 400065, China (e-mail:liaoxi@cqupt.edu.cn).}
\thanks{Corresponding author: Yinsheng Liu.}
%\thanks{This work was supported by Fundamental Research Funds for the Central Universities (2020JBM085, 2020JBZD005), State Key Laboratory of Rail Traffic Control and Safety (Contract No.RCS2020ZT009) Beijing Jiaotong University, National Key R\&D Program of China under Grant 2020YFB1804901, and Chongqing Municipal Key Laboratory of Institutions of Higher Education. }

}
\maketitle
%\doublespacing
\begin{abstract}
%\boldmath
Spatial covariance matrix (SCM) is essential in many applications of multi-antenna systems such as massive multiple-input multiple-output (MIMO). For massive MIMO operating at millimeter-wave bands, hybrid analog-digital structure has been adopted to reduce the cost of radio frequency (RF) chains. In this situation, signals received at the antennas are unavailable to the digital receiver, and as a consequence, traditional sample average approach cannot be used for SCM reconstruction in hybrid massive MIMO. To address this issue, beam sweeping algorithm (BSA), which can reconstruct SCM effectively in hybrid massive MIMO, has been proposed in our previous work. In this paper, a truncated BSA is further proposed for SCM reconstruction by taking into account the patterns of antenna elements in the array. Due to the directive antenna pattern, sweeping results corresponding to predetermined direction-of-angles (DOA) far from the normal direction are small and thus can be replaced by predetermined constants. At the cost of negligible performance reduction, SCM can be reconstructed efficiently by sweeping only the predetermined DOAs that are close to the normal direction. In this way, BSA can be conducted much faster than its traditional counterpart. Insightful analysis will be also included to show the impact of truncation on the performance.
\end{abstract}


%\newpage
\section{Introduction}
Spatial covariance matrix (SCM) is essential in many applications of multi-antenna systems \cite{TETuncer,ROSchmidt2}, such as massive multiple-input multiple-output (MIMO). Massive MIMO is one of the most important enabling technologies in 5G and its beyond \cite{EGLarsson}. Due to a large number of antennas, massive MIMO is essential to millimeter-wave bands because the large array gain can compensate for the high path loss, and frequency resources at millimeter-wave bands can be therefore exploited efficiently \cite{LLiang,LYou}.\par

To reduce the number of radio frequency (RF) chains, hybrid structure has been adopted for massive MIMO operating at millimeter-wave bands \cite{OEAyach1,VVen,CLin}. In hybrid systems, one RF chain is connected to multiple antennas, so that the number of RF chains can be greatly reduced. However, in hybrid massive MIMO, the received signals at the antennas are first fed to the analog phase shifters and then combined in the analog domain before sent to the digital receiver. Consequently, the received signals at the antennas are unavailable to the digital receiver, and thus traditional sample average approach cannot be used for SCM reconstruction \cite{GMan}. To address this issue, we have developed a beam sweeping algorithm (BSA) for SCM reconstruction in hybrid massive MIMO systems \cite{SLi}. In this approach, beam sweeping results corresponding to a group of predetermined direction-of-angles (DOA) are collected and then SCM can be reconstructed by solving a matrix equation. To reduce the complexity caused by solving matrix equation in \cite{SLi}, a high-efficiency BSA is proposed in \cite{YZhou} through apropriately adjusting the weights connected to each antenna. Also, predetermined DOAs are carefully selected in \cite{YLiu} so that matrix inverse can be completely avoided. In addition, BSA in \cite{YLiu} is also improved so that it can be used in the presence of multiple RF chains under the framework of sub-connected massive MIMO \cite{CLin}. Moreover, the issue of high computational complexity can be also addressed by adopting new computation platform, such as quantum computer \cite{FMeng}.

The weakness of BSA lies in that it needs a large number of beam sweeping operations, and thus the algorithm delay caused by beam sweeping can be significant. Although fully-connected hybrid structure can be used to reduce algorithm delay \cite{ZFU}, that approach is unavailable in the cases of single RF chain or sub-connected hybrid structure. To address this issue, we propose a truncated BSA in this paper. In this approach, patterns of antenna elements in the array are exploited. Due to the weak response of the antenna pattern at the DOAs far from the normal direction, the received power from those DOAs can be very small. Therefore, sweeping results corresponding to predetermined DOAs far from the normal direction can be truncated. Then, SCM can be efficiently reconstructed by sweeping only the predetermined DOAs that are close to the normal direction at the cost of negligible performance reduction. In this way, BSA can be conducted much faster than its traditional counterpart in \cite{SLi,YLiu}. Insightful discussion will be also presented to show the impact of truncation on the reconstruction accuracy, as well as a theoretical analysis on the asymptotical results which are not revealed in \cite{SLi,YLiu}.

The rest of this paper is organized as follows. In Section II, signal model for hybrid massive MIMO is introduced, followed by a review of SCM reconstruction issue. In Section III, truncated BSA is presented, and insights will be shown in Section IV. Simulation results and conclusions are in Section V and VI, respectively.

\section{System Model}

\subsection{Signal Model}
As in Fig.~\ref{system}, consider a hybrid massive MIMO receiver composed of a uniform linear array (ULA) with $M$ antennas. To simplify the symbol notation, a single RF chain is considered in this paper even though the proposed approach can be also used in the case of multiple RF chains, as in \cite{YLiu}.\par

For a general model, patterns of antenna elements in the array should be taken into account. Since all elements in the ULA are the same, all antenna elements can share a common antenna pattern $g(\theta)$ where $\theta$ indicates the DOA. Therefore, if denote $y_{m}(t)$ to be the received signal on the $m$-th antenna with $m=0,1,\cdots,M-1$, then the received signal vector  $\boldsymbol{y}(t)=[y_0(t),y_1(t),\cdots,y_{M-1}(t)]^{\mathrm{T}}$ can be represented as
\begin{align}\label{signalmodel}
\boldsymbol{y}(t)=\sum_{l=0}^{L-1}\boldsymbol{a}(\theta_l)g(\theta_l)x_l(t)+\boldsymbol{z}(t),
\end{align}
where $x_l(t)$'s ($l=0,1,\cdots,L - 1)$ are $L$ signals impinging from far field onto the array, $\theta_l$ is the DOA of $x_l(t)$, and $\boldsymbol{z}(t)$ denotes the additive Gaussian noise vector with $\mathrm{E}\{\boldsymbol{z}(t)\boldsymbol{z}^{\mathrm{H}}(t)\}=N_0\boldsymbol{I}$ with $N_0$ and $\boldsymbol{I}$ being the noise power and an identity matrix, respectively. In (\ref{signalmodel}), $\boldsymbol{a}(\theta_l)$ is the $M\times 1$ steering vector with the $m$-th entry given by
$
a_m(\theta_l)=e^{j2\pi\cdot\frac{d}{\lambda}\cdot\sin\theta_l\cdot m},
$
where $d=0.5\lambda$ is the antenna distance and $\lambda$ is the wave length. If assuming $L$ signals are mutually independent with zero means and the power of the $l$-th signal is $\mathrm{E}\{|x_l(t)|^2\}=\sigma_l^2$, then the SCM, $\boldsymbol{R}=\mathrm{E}\{\boldsymbol{y}(t)\boldsymbol{y}^{\mathrm{H}}(t)\}$, can be obtained as
\begin{align}\label{R}
\boldsymbol{R}=\sum_{l=0}^{L-1}\sigma_l^2|g(\theta_l)|^2\boldsymbol{a}(\theta_l)\boldsymbol{a}^{\mathrm{H}}(\theta_l) + N_0\boldsymbol{I}.
\end{align}

Note that the signal model in (\ref{signalmodel}) can be also used even in the presence of mutual coupling among antennas. As in \cite{WLStutzman}, most antenna elements in a long ULA can see the same neighboring environments, and therefore an average antenna pattern, which is common to all antennas, can be employed to describe the mutual coupling effect. However, due to mutual coupling, the average antenna pattern $g(\theta)$ may deviate from the one designed for the single antenna case. Active measurement can be used in this case to figure out the practical $g(\theta)$.

\subsection{SCM Reconstruction}
\begin{figure}
  \centering
  \includegraphics[width=3.5in]{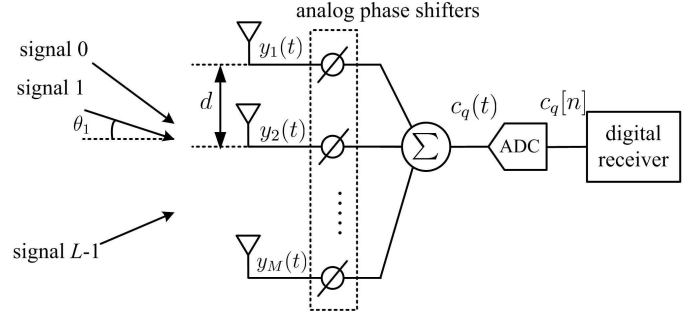}\\
  \caption{Hybrid massive MIMO receiver with a single RF chain and a ULA.}\label{system}
\end{figure}

Denote $\boldsymbol{y}[n]=\boldsymbol{y}(nT_s)$ to be the sample of received signal where $T_s$ denotes the sampling period. If all entries of $\boldsymbol{y}[n]$ are available to the digital receiver, then SCM in (\ref{R}) can be estimated using the sample average approach, that is \cite{GMan}
\begin{align}\label{approx}
\widehat{\boldsymbol{R}}= \frac{1}{N}\sum_{n=0}^{N-1}\boldsymbol{y}[n]\boldsymbol{y}^{\mathrm{H}}[n],
\end{align}
where $N$ denotes the number of samples. In this approach, received signals at all antennas should be sent via RF chains to the digital receiver. In hybrid massive MIMO, however, Fig.~\ref{system} shows that only the combination of the entries of $\boldsymbol{y}[n]$ can be seen by the digital receiver because there is only one RF chain. As a consequence, the sample average approach in (\ref{approx}) cannot be used in hybrid massive MIMO systems.\par

To address this issue, we have developed basic BSA for SCM reconstruction in hybrid massive MIMO with single RF chain \cite{SLi}. Then, the basic BSA has been improved in \cite{YLiu} to handle the case of multiple RF chains. A careful selection of predetermined DOAs has also been presented in \cite{YLiu} so that the computational complexity can be greatly reduced.

In spite of the success in SCM reconstruction, the weakness of BSA is obvious because it needs a large number of beam sweeping operations and the algorithm delay caused by beam sweeping can be significant. Although fully-connected hybrid structure can be used to reduce algorithm delay \cite{ZFU}, that approach is unavailable in the cases of single RF chain or sub-connected hybrid structure. To overcome this issue, a truncated BSA is proposed as we will show in Section III.

\section{Truncated BSA}
As the truncated BSA relies on the traditional BSA, we will first review BSA in \cite{SLi} and \cite{YLiu} in brief to provide a framework for the truncated BSA.
\subsection{Review on BSA}
As in \cite{SLi}, define $\{\theta^{(0)},\theta^{(1)},\cdots,\theta^{(Q-1)}\}$ to be a set of predetermined DOAs. The analog beamformers switch the beam directions to the predetermined DOAs in turn. For the $q$-th beam, the combination of the received signals can be represented by $c_q(t)=\boldsymbol{a}^{\mathrm{H}}(\theta^{(q)})\boldsymbol{y}(t)$. From Fig.~\ref{system}, the signal combination is sampled before send to the receiver, and thus the samples of the signal combination can be given by
\begin{align}
c_q[n]=c_q(nT_s)=\boldsymbol{a}^{\mathrm{H}}(\theta^{(q)})\boldsymbol{y}[n].
\end{align}
Denote $P_q=\mathrm{E}(|c_q[n]|^2)$ to be the statistical power of $c_q[n]$. Using sample average, an estimation of $P_q$ can be obtained as
\begin{align}\label{sampleave}
\widehat{P}_q=\frac{1}{N}\sum_{n=0}^{N-1}|c_q[n]|^2=\boldsymbol{a}^{\mathrm{H}}(\theta^{(q)})\widehat{\boldsymbol{R}}\boldsymbol{a}(\theta^{(q)}).
\end{align}
Using $\mathrm{vec}(\cdot)$ operator to (\ref{sampleave}), $\widehat{P}_q$ can be rewritten as $\widehat{P}_q=\boldsymbol{a}_q^{\mathrm{T}}\widehat{\boldsymbol{r}}$ where $\widehat{\boldsymbol{r}}=\mathrm{vec}(\widehat{\boldsymbol{R}})$ and $\boldsymbol{a}_q=\boldsymbol{a}(\theta^{(q)})\otimes\boldsymbol{a}^*(\theta^{(q)})$ with $\otimes$ indicating Kronecker product. If taking all $Q$ predetermined DOAs into account, we can derive that
\begin{align}\label{equ}
\boldsymbol{A}\widehat{\boldsymbol{r}}=\widehat{\boldsymbol{p}},
\end{align}
where $\widehat{\boldsymbol{p}}=(\widehat{P}_0,\widehat{P}_1,\cdots,\widehat{P}_{Q-1})^{\mathrm{T}}$ contains the estimated power on all predetermined beams and $\boldsymbol{A}=(\boldsymbol{a}_0,\boldsymbol{a}_1,\cdots,\boldsymbol{a}_{Q-1})^{\mathrm{T}}$. As in \cite{SLi}, unknown $\widehat{\boldsymbol{r}}$ can be obtained by solving (\ref{equ}) so that the desired SCM can be reconstructed as $\widehat{\boldsymbol{R}}=\mathrm{unvec}(\widehat{\boldsymbol{r}})$.\par

Direct solution of (\ref{equ}) suffers from high complexity due to matrix inverse. To avoid this issue, a low-complexity BSA is investigated in \cite{YLiu}. In this approach, (\ref{equ}) is converted to
\begin{align}\label{rbr}
\boldsymbol{B}\widehat{\boldsymbol{\gamma}}=\widehat{\boldsymbol{p}},
\end{align}
where $\boldsymbol{B}=(\boldsymbol{b}_0,\boldsymbol{b}_1,\cdots,\boldsymbol{b}_{Q-1})^{\mathrm{T}}$ with $\boldsymbol{b}_q$ being a $(2M-1)\times 1$ vector given by
\begin{align}
\boldsymbol{b}_q=
\left(
  \begin{array}{ccc}
    a_{M-1}(\theta^{(q)}) & \cdots & 0 \\
    \vdots &  & \vdots \\
    a_0(\theta^{(q)}) & \cdots & 0 \\
    0 & \cdots & a_{M-1}(\theta^{(q)}) \\
    \vdots &  & \vdots \\
    0 & \cdots & a_0(\theta^{(q)}) \\
  \end{array}
\right)\boldsymbol{a}^*(\theta^{(q)}),
\end{align}
and $\widehat{\boldsymbol{\gamma}}=(\widehat{\gamma}[1-M],\cdots,\widehat{\gamma}[M-1])^{\mathrm{T}}$ corresponds to $2M-1$ non-repeated entries in $\boldsymbol{R}$ with $\widehat{\gamma}[m-n]$ being the $(m,n)$-th entry of $\widehat{\boldsymbol{R}}$, that is, $\widehat{\gamma}[m-n]=[\widehat{\boldsymbol{R}}]_{m,n}$. If the predetermined DOAs are selected as $\theta^{(q)}=2\mathrm{arcsin}(-1+2q/Q)$, then the number of predetermined DOAs can be determined as $Q=2M-1$ and $\boldsymbol{B}^{\mathrm{H}}\boldsymbol{B}$ becomes a diagonal matrix $\boldsymbol{\Delta}$ with the $m$-th $(m=0,1,\cdots,2M-2)$ diagonal entry given by
\begin{align}\label{Delta}
[\boldsymbol{\Delta}]_{m,m}=\begin{cases}
Q(m + 1), &m\leq M-1\\
Q(2M - 1 - m), &m> M-1
\end{cases}.
\end{align}
Therefore, unknown $\widehat{\boldsymbol{\gamma}}$ can be obtained as
\begin{align}\label{recon}
\widehat{\boldsymbol{\gamma}}=\boldsymbol{\Delta}^{-1}\boldsymbol{B}^{\mathrm{H}}\widehat{\boldsymbol{p}},
\end{align}
where matrix inverse has been avoided. The desired SCM can be then reconstructed using the estimations of the entries contained in $\widehat{\boldsymbol{\gamma}}$.
\subsection{Truncated Beam Sweeping}
In \cite{SLi} and \cite{YLiu}, omni-directional antenna elements are implicitly assumed, and therefore full beam sweeping has to be adopted where power estimation in (\ref{sampleave}) is conducted for each predetermined DOA. Actually, full beam sweeping can be approximated by a truncated beam sweeping if we exploit the antenna pattern $g(\theta)$.\par
To this end, we assume the number of samples is large enough. Then, the sample average in (\ref{sampleave}) can be replaced by the statistical average and thus
\begin{align}\label{tight}
\widehat{P}_q&=P_q=\boldsymbol{a}^{\mathrm{H}}(\theta^{(q)})\boldsymbol{R}\boldsymbol{a}(\theta^{(q)}).
\end{align}
Substitute (\ref{R}) into (\ref{tight}), we have
\begin{align}\label{insights}
P_q&=\sum_{l=0}^{L-1}\sigma_l^2|g(\theta_l)|^2|\boldsymbol{a}^{\mathrm{H}}(\theta^{(q)})\boldsymbol{a}(\theta_l)|^2+MN_0\nonumber\\
&=\sum_{l=0}^{L-1}\sigma_l^2M^2|g(\theta_l)|^2\left|\mathrm{sinc}\left[\pi\frac{d}{\lambda}M\left(\sin\theta_l-\sin\theta^{(q)}\right)\right]\right|^2\nonumber\\
&~~~~~+MN_0.
\end{align}
When the number of antennas is large, the feature of $\mathrm{sinc}(\cdot)$ function in (\ref{insights}) indicates that only arriving signals with DOAs close to $\theta^{(q)}$ can cause significant contribution to $P_q$. Therefore, if the antenna pattern $g(\theta)$ has very weak response for DOAs that are far from the normal direction, the summarization term in (\ref{insights}) can be approximated by zero when $\theta^{(q)}$ is far from the normal direction. The mathematical observation in above also coincides with the following physical intuition: as the overall array response $g(\theta)\boldsymbol{a}(\theta)$ is scaled by $g(\theta)$, the received power could be very weak if we point the beam to a direction that is far from the normal direction.\par

Inspired by the above observation, the full beam sweeping in (\ref{sampleave}) can be approximated by the following truncated beam sweeping operation
\begin{align}\label{trunc}
\widehat{P}_q=
\begin{cases}
\displaystyle{\frac{1}{N}\sum_{n=0}^{N-1}|c_q[n]|^2},&q\in\mathcal{S}\\
~~~~~MN_0,&q\in\overline{\mathcal{S}}
\end{cases}.
\end{align}
In (\ref{trunc}), $\mathcal{S}=\{q|\left\lceil T/2\right\rceil\leq q< Q-\lfloor T/2\rfloor\}$ where $T$ is an integer ($0\leq T\leq Q-1$) indicating the number of truncated predetermined DOAs with $\left\lceil\cdot\right\rceil$ and $\left\lfloor\cdot\right\rfloor$ indicating round up and down to the nearest integers, and $\overline{\mathcal{S}}$ is the complementary set of ${\mathcal{S}}$. It means in (\ref{trunc}) that power estimation is only need for $\theta^{(q)}$'s that are close to the normal direction, while a predetermined constant can be used instead for $\theta^{(q)}$'s that are far from the normal direction. Since only $Q-T$ predetermined DOAs needs power estimation, truncated beam sweeping operation can be accelerated when $T>0$, at the cost of negligible performance reduction as will be shown in Section IV.\par

Once the truncated sweeping results in (\ref{trunc}) are available, they can be used in (\ref{recon}) to reconstruct the desired SCM.
\section{Insights}
To derive insightful results in this section, we assume that $\theta_l\in\{\theta^{(0)},\theta^{(1)},\cdots,\theta^{(Q-1)}\}$, and thus the DOA associated with the $l$-th signal can be rewritten as $\theta_l=\theta^{(q_l)}$. In this case, the $\mathrm{sinc}(\cdot)$ function in (\ref{insights}) can be approximated by
\begin{align}\label{assume}
\mathrm{sinc}\left[\pi\frac{d}{\lambda}M\left(\sin\theta^{(q_l)}-\sin\theta^{(q)}\right)\right]\approx \delta[q_l-q],
\end{align}
where $\delta[\cdot]$ indicates Kronecker-delta function.
\subsection{Performance Reduction}
To evaluate the performance reduction caused by truncation, define $\|\widehat{\boldsymbol{R}}-\boldsymbol{R}\|_{\mathrm{F}}^2$ to be the squared-error (SE) between desired SCM and reconstructed SCM. Then, we can obtain
\begin{align}\label{Fnorm}
\|\widehat{\boldsymbol{R}}-\boldsymbol{R}\|_{\mathrm{F}}^2=\sum_{m=0}^{M-1}\sum_{n=0}^{M-1}|\widehat{\gamma}[m-n]-\gamma[m-n]|^2,
\end{align}
where $\gamma[m-n]=[\boldsymbol{R}]_{m,n}$.
Recalling that $\boldsymbol{R}$ have $2M-1$ non-repeated entries in total, (\ref{Fnorm}) can be rewritten as
\begin{align}\label{Fnorm1}
\|\widehat{\boldsymbol{R}}-\boldsymbol{R}\|_{\mathrm{F}}^2&=\sum_{m=1-M}^{M-1}(M-|m|)\cdot|\widehat{\gamma}[m]-\gamma[m]|^2\nonumber\\
&=Q^{-1}\|\boldsymbol{\Delta}^{1/2}(\widehat{\boldsymbol{\gamma}}-\boldsymbol{\gamma})\|_{2}^2,
\end{align}
where we have used the identity in (\ref{Delta}) for the second equation. In (\ref{Fnorm1}), $\boldsymbol{\gamma}=({\gamma}[1-M],\cdots,{\gamma}[M-1])^{\mathrm{T}}$ contains $2M-1$ non-repeated entries in $\boldsymbol{R}$. If denote $\boldsymbol{p}=({P}_0,{P}_1,\cdots,{P}_{Q-1})^{\mathrm{T}}$, then we can derive, similar to (\ref{rbr}) and (\ref{recon}), that $\boldsymbol{B}{\boldsymbol{\gamma}}={\boldsymbol{p}}$ and ${\boldsymbol{\gamma}}=\boldsymbol{\Delta}^{-1}\boldsymbol{B}^{\mathrm{H}}{\boldsymbol{p}}$. Using this identities and (\ref{recon}), (\ref{Fnorm1}) can be rewritten as
\begin{align}
\|\widehat{\boldsymbol{R}}-\boldsymbol{R}\|_{\mathrm{F}}^2=Q^{-1}(\widehat{\boldsymbol{p}}-\boldsymbol{p})^{\mathrm{H}}\boldsymbol{B}\boldsymbol{\Delta}^{-1}\boldsymbol{B}^{\mathrm{H}}(\widehat{\boldsymbol{p}}-\boldsymbol{p}).
\end{align}
To simplify the expression $\boldsymbol{B}\boldsymbol{\Delta}^{-1}\boldsymbol{B}^{\mathrm{H}}$, consider that
\begin{align}\label{trans}
\boldsymbol{B}\boldsymbol{\Delta}^{-1}\boldsymbol{\Delta}=\boldsymbol{B}.
\end{align}
Since $\boldsymbol{B}^{\mathrm{H}}\boldsymbol{B}=\boldsymbol{\Delta}$, (\ref{trans}) can be rewritten as $\boldsymbol{B}\boldsymbol{\Delta}^{-1}\boldsymbol{B}^{\mathrm{H}}\boldsymbol{B}=\boldsymbol{B}$ or equivalently
\begin{align}\label{result}
(\boldsymbol{B}\boldsymbol{\Delta}^{-1}\boldsymbol{B}^{\mathrm{H}}-\boldsymbol{I})\boldsymbol{B}=\boldsymbol{O},
\end{align}
where $\boldsymbol{O}$ is a $(2M-1)\times (2M-1)$ all zero matrix. Recalling that $2M-1=\mathrm{Rank}(\boldsymbol{B}^{\mathrm{H}}\boldsymbol{B})\leq \mathrm{Rank}(\boldsymbol{B})\leq 2M-1$, $\boldsymbol{B}$ is therefore a non-singular matrix and equation $\boldsymbol{X}\boldsymbol{B}=\boldsymbol{O}$ has only one solution $\boldsymbol{X}=\boldsymbol{O}$. Therefore, it shows in (\ref{result}) that
\begin{align}
\boldsymbol{B}\boldsymbol{\Delta}^{-1}\boldsymbol{B}^{\mathrm{H}}=\boldsymbol{I}.
\end{align}
As a result, we have
\begin{align}\label{20}
\|\widehat{\boldsymbol{R}}-\boldsymbol{R}\|_{\mathrm{F}}^2=Q^{-1}\|\widehat{\boldsymbol{p}}-\boldsymbol{p}\|_2^2.
\end{align}
Then, using (\ref{insights}) and the assumption in (\ref{assume}), (\ref{20}) can be obtained as
\begin{align}\label{inn}
\|\widehat{\boldsymbol{R}}-\boldsymbol{R}\|_{\mathrm{F}}^2=\frac{M^2}{Q}\sum_{q\in\overline{\mathcal{S}}}\sum_{l=0}^{L-1}\sigma_l^2|g(\theta^{(q_l)})|^2\delta[q_l-q].
\end{align}

It shows in (\ref{inn}) that the $l$-th received signal can contribute to the SE only when $q_l\in\overline{\mathcal{S}}$. As a result, performance reduction due to truncated beam sweeping can be ignored because $g(\theta^{(q_l)})$ is very small if $q_l\in\overline{\mathcal{S}}$.
\subsection{Asymptotical Analysis}
To evaluate the impact of antenna patterns on the performance, asymptotical analysis is used. In this case, the number of samples is finite, and thus mean-SE (MSE) should be used instead. $T=0$ is considered so that no truncation is employed. In addition to the assumption in (\ref{assume}), we also assume that $x_l[n]=x_l(nT_s)\sim\mathrm{CN}(0,\sigma_l^2)$.\par
Under this situations, from (\ref{20}), we can obtain
\begin{align}\label{mse}
\mathrm{E}(\|\widehat{\boldsymbol{R}}-\boldsymbol{R}\|_{\mathrm{F}}^2)=Q^{-1}\sum_{q=0}^{Q-1}\mathrm{E}(|\widehat{P}_q-P_q|^2).
\end{align}
Since $\mathrm{E}(\widehat{\boldsymbol{R}})=\boldsymbol{R}$, we have
\begin{align}\label{error}
\mathrm{E}(|\widehat{P}_q-P_q|^2)&=\mathrm{E}[|\boldsymbol{a}^{\mathrm{H}}(\theta^{(q)})\widehat{\boldsymbol{R}}\boldsymbol{a}(\theta^{(q)})|^2]\nonumber\\
&~~~~-|\boldsymbol{a}^{\mathrm{H}}(\theta^{(q)}){\boldsymbol{R}}\boldsymbol{a}(\theta^{(q)})|^2.
\end{align}
Then, substitute (\ref{signalmodel}) into the first term in the right-side of (\ref{error}), and after a long and tedious mathematical procedure, we obtain
\begin{align}\label{sympt}
&\mathrm{E}(|\widehat{P}_q-P_q|^2)=\frac{1}{N}\left(\sum_{l=0}^{L-1}|\boldsymbol{a}^{\mathrm{H}}(\theta^{(q)})\boldsymbol{a}(\theta_l)g(\theta_l)|^2\sigma_l^2\right)^2+\nonumber\\
&~~\frac{2M N_0}{N}\sum_{l=0}^{M-1}|\boldsymbol{a}^{\mathrm{H}}(\theta^{(q)})\boldsymbol{a}(\theta_l)g(\theta_l)|^2\sigma_l^2+\frac{M^2N_0^2}{N},
\end{align}
where we have used the identity $\mathrm{E}(x_1x_2x_3x_4)=\mathrm{E}(x_1x_2)\mathrm{E}(x_3x_4)+\mathrm{E}(x_1x_3)\mathrm{E}(x_2x_4)+\mathrm{E}(x_1x_4)\mathrm{E}(x_2x_3)$ for general Gaussian random variables $x_1,x_2,x_3,x_4$\cite{CLNikias}.

Then, with the assumption in (\ref{assume}), (\ref{sympt}) is simplified as
\begin{align}
\mathrm{E}(|\widehat{P}_q-P_q|^2)=\frac{M^4}{N}\sum_{l=0}^{L-1}\sigma_l^4|g(\theta^{(q_l)})|^4\delta[q-q_l]+\nonumber\\
\frac{2M^3N_0}{N}\sum_{l=0}^{L-1}\sigma_l^2|g(\theta^{(q_l)})|^2\delta[q-q_l]+\frac{M^2N_0^2}{N}.
\end{align}
The MSE in (\ref{mse}) can be therefore obtained as
\begin{align}\label{27}
\mathrm{E}(\|\widehat{\boldsymbol{R}}-\boldsymbol{R}\|_{\mathrm{F}}^2)=\frac{M^4}{QN}\sum_{l=0}^{L-1}|g(\theta^{(q_l)})|^4\sigma_l^4+\nonumber\\
\frac{2M^3N_0}{QN}\sum_{l=0}^{L-1}|g(\theta^{(q_l)})|^2\sigma_l^2+\frac{M^2N_0^2}{QN}.
\end{align}

Following insights can be obtained from (\ref{27}):
Fist, due to the antenna pattern $g(\theta)$, arriving signals with DOAs close to the normal direction contribute more to the MSE than the ones with DOAs far from the normal direction. This observation can justify the truncated BSA because lost of sweeping results at predetermined DOAs far from normal direction has little impact on the performance. Second, the MSE can vanish as the increasing of $N$, which coincides with the intuition and the simulation results in \cite{SLi} and \cite{YLiu}.

\section{Simulation}
Computer simulation is adopted in this section to demonstrate the truncated BSA. Consider a ULA with $M=16$ antennas, and the distance between neighboring antennas is $0.5\lambda$. Two signals arrives at the ULA. DOAs associated with the first and the second signals are $0^\mathrm{o}$ and $0<\theta<90^{\mathrm{o}}$, respectively. Arriving signals are assumed independent with zero means and unit powers, and the signal-to-noise ratio is $0$ dB. The number of predetermined DOAs is $Q=2M-1$ and the predetermined DOAs are selected as $\theta^{(q)}=2\mathrm{arcsin}(-1+2q/Q)$. Similar to \cite{SLi} and \cite{YLiu}, normalized SE (NSE) is used to evaluate the accuracy of reconstructed SCM, that is
$
\mathrm{NSE}=\|\widehat{\boldsymbol{R}}-\boldsymbol{R}\|_{\mathrm{F}}^2\cdot\|\boldsymbol{R}\|_{\mathrm{F}}^{-2}.
$
To take different antenna patterns into account, the following pattern function is employed
$
g(\theta)=\mathrm{sinc}(\alpha\sin\theta),
$
where $\alpha$ is a constant that can adjust the pattern of $g(\theta)$. The number of samples is $N=5000$, and the simulation results are averaged over $10000$ runs.

\begin{figure}
  \centering
  \includegraphics[width=3.6in]{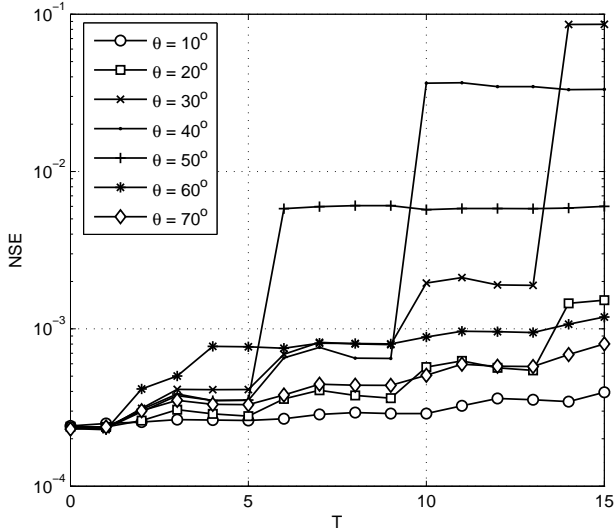}\\
  \caption{NSE versus the number of truncated predetermined DOAs with different DOAs for the second arriving signal. In this figure, $\alpha=1$.}\label{fig1}
\end{figure}

NSE versus the number of truncated predetermined DOAs are shown in Fig.~\ref{fig1} with $\alpha=1$. Generally, NSE arises as the increasing of $T$ because more predetermined beams are truncated. When the DOA associated with the second arriving signal is small (e.g. $\theta=10^{\mathrm{o}},20^{\mathrm{o}}$), performance reduction due to the rising of $T$ can be ignored because the information of spatial distribution for arriving signals has been captured even though a half of predetermined beams ($T_{\mathrm{max}}/Q\approx 0.5$) are truncated. For larger DOAs (e.g. $\theta=30^{\mathrm{o}},40^{\mathrm{o}},50^{\mathrm{o}}$), a significant performance degradation can be observed if $T$ is large enough. When $T$ is large, predetermined DOAs far from the normal direction cannot be swept and thus a significant degradation can be observed. However, as the further increasing of $\theta$ (e.g. $\theta=60^{\mathrm{o}},70^{\mathrm{o}}$), the performance reduction becomes small again. This is because lost of sweeping results at $\theta=60^{\mathrm{o}},70^{\mathrm{o}}$ causes little impact on the performance since the antenna pattern $g(\theta)$ has very weak response at these directions.

\begin{figure}
  \centering
  \includegraphics[width=3.6in]{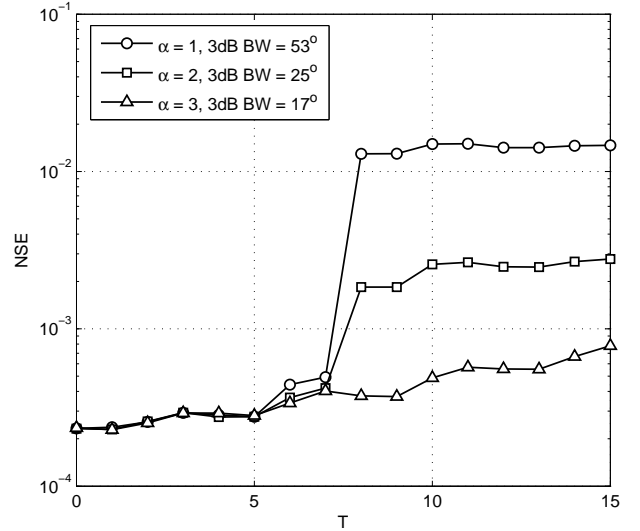}\\
  \caption{NSE versus the number of truncated predetermined DOAs with different antenna patterns. In this figure, $\theta=45^{\mathrm{o}}$.}\label{fig2}
\end{figure}

Fig.~\ref{fig2} shows the impact of different antenna patterns on the reconstruction accuracy. The antenna patterns are adjusted by setting $\alpha=1,2,3$ respectively. The corresponding beam widths (BW) of mainlobe are $53^{\mathrm{o}},25^{\mathrm{o}},17^{\mathrm{o}}$. When $\alpha$ is small, the main beam of $g(\theta)$ is wide and the arriving signal at $\theta=45^{\mathrm{o}}$ plays an important role in the SCM. As a result, when $T$ is large enough, information on the spatial distribution cannot be captured and thus a significant performance degradation can be observed. When $\alpha$ is large, the main beam of $g(\theta)$ is narrow, and thus the arriving signal at $\theta=45^{\mathrm{o}}$ contribute little to the SCM. Therefore, lost of information on spatial distribution corresponding to $\theta=45^{\mathrm{o}}$ causes little performance reduction.

\section{Conclusions}
In order to reduce the algorithm delay caused by full beam sweeping, we have proposed a truncated BSA in this paper. By exploiting the feature of antenna patterns, the predetermined DOAs that are far from the normal direction can be replaced by a constant and therefore power estimation on corresponding beams are not required any more. In this way, BSA approach can be accelerated at the cost of  negligible performance reduction. Insightful analysis, including the performance degradation caused by truncation and the asymptotically analysis on the effect of antenna pattern, have also been presented in this paper. Simulation results have also been presented to justify the truncated BSA.

\bibliographystyle{IEEEtran}
\bibliography{IEEEabrv,MUSIC}

\end{document}